\documentclass[aps,pre,floats,twocolumn,amsfonts,amssymb,unsortedaddress,floatfix]{revtex4}
\usepackage{epsfig}
\usepackage{amssymb}
\usepackage{mathtools}
\usepackage{color}

\usepackage{tikz}

\newcommand{\tikzcircle}[2][black,fill=black]{\tikz[baseline=-0.5ex]\draw[#1,radius=#2] (0,0) circle ;}%
\newcommand{\tikzcirclew}[2][black,fill=white]{\tikz[baseline=-0.5ex]\draw[#1,radius=#2] (0,0) circle ;}%
\newcommand{\tikzcircler}[2][red,fill=white]{\tikz[baseline=-0.5ex]\draw[#1,radius=#2] (0,0) circle ;}

\begin{document}

\title{Unexpected Phenomenology in Particle-Based Ice Absent in Magnetic Spin Ice} 

\author{Cristiano Nisoli}

\affiliation{Theoretical Division, Los Alamos National Laboratory, Los Alamos, NM, 87545, USA}\email[]{cristiano.nisoli@gmail.com, cristiano@lanl.gov}

\date{\today}

\begin{abstract}
While particle-based ices are often considered essentially equivalent to magnet-based spin ices, the two
differ essentially in frustration and energetics. We show that at equilibrium particle-based ices correspond
exactly to spin ices coupled to a background field. In trivial geometries, such a field has no effect, and the
two systems are indeed thermodynamically equivalent. In other cases, however, the field controls a richer
phenomenology, absent in magnetic ices, and still largely unexplored: ice rule fragility, topological charge
transfer, radial polarization, decimation induced disorder, and glassiness.
\end{abstract}

\maketitle

{\it Introduction.} The ice rule~\cite{bernal1933theory} has had an impactful history.  Pauling employed it to explain~\cite{pauling1935structure} the zero point entropy of water ice~\cite{giauque1936entropy} as a consequence of the degeneracy in allocating two protons close to, and two away from, each oxygen atom sitting in any of the tetrahedron-based  crystal structures of ice. However, the concept is   more general. Consider  binary spins placed along the edges of a graph, impinging in its vertices (Fig.~1). The topological charge of  a vertex of coordination $z$ is the difference between the $n$ spins pointing in and the $z-n$ pointing out, or $q_n=2n-z$ (Fig 1). Then, an ice-manifold is the degenerate set of spin configurations that minimizes $|q|$ locally. If $z$ is even, the minimal $|q|$ is zero (for $z=4$, we recover the original ice rule, 2-in/2-out, of water ice and rare earth titanates magnets~\cite{Ramirez1999}). If $z$ is  odd, ice rule vertices have charges $q=\pm1$ and the ice-manifold is a neutral plasma of topological charges~\cite{Moller2009, Chern2011,Rougemaille2011,Zhang2013,anghinolfi2015thermodynamic}.

As Ice manifolds  can typically host unusual phases~\cite{nisoli2017deliberate}, they have  invited  the design of a new class of artificial, frustrated magnetic nano-materials, called ``artificial spin ices'' (SI). These are arrays of interacting, single-domain, shape-anisotropic, magnetic nano-islands whose magnetizations are described by binary spin and obey the ice rule (Fig.~1a,b) \cite{Wang2006,Nisoli2013colloquium}. Their  exotic behaviors are often not found in natural magnets~\cite{Morrison2013} and  can be designed to study memory effects~\cite{gilbert2015direct}, effective thermodynamics in driven systems~\cite{Nisoli2010}, magnetic charges and monopoles~\cite{Mengotti2010,rougemaille2011artificial,ladak2011direct,Zhang2013}, anomalous hall effects~\cite{Branford2012,PhysRevB.95.060405}, often with real time, real space characterization~\cite{Farhan2013,kapaklis2014thermal,Porro2013,gilbert2016emergent,gilbert2016frustration}.

``Particle ices'' (PI) are another artificial implementation of an ice manifold~\cite{Libal2006,libal2009,Olson2012,libal2017inner,libal2012hysteresis,libal2017dynamic}. Mutually repulsive particles are trapped, one particle per trap, with preferential occupation at its extremes (Fig.~1c,d). Traps are arranged along the edges of a lattice whose geometry determines the collective behavior. They have been studied numerically~\cite{Libal2006,libal2009,Olson2012} and  realized experimentally using magnetic colloids gravitationally trapped in microgrooves~\cite{ortiz2016engineering,loehr2016defect} but also in flux quanta pinned to nano-patterned superconductors~\cite{Latimer2013,Trastoy2014freezing,ge2017direct}.  As PI was also found to obey the ice rule, at least in the square and hexagonal geometry, ideas and results have been exchanged among PI and SI, often considered as essentially equivalent systems. That assumption is incorrect. 

\begin{figure}[b!]
\center
\includegraphics[width=1\columnwidth]{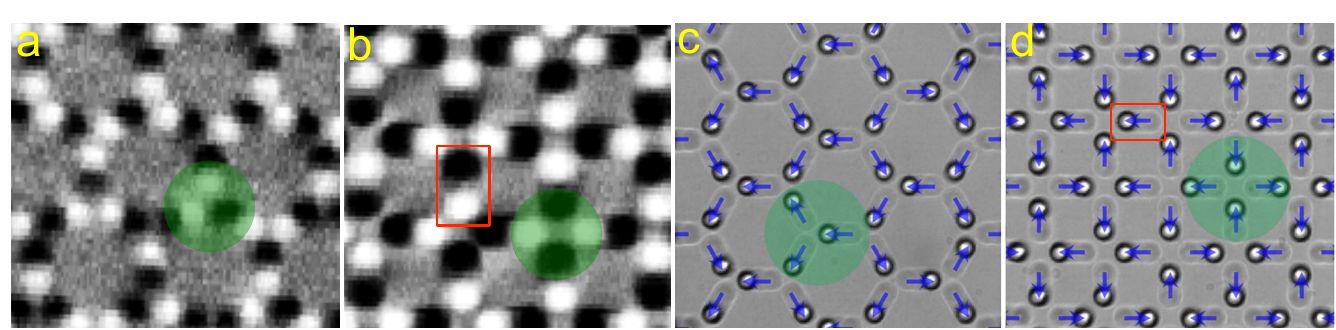}
\begin{center}
\caption{Magnetic force microscopy of hexagonal (a) and square (b) SI show  the  constitutive degrees of freedom (red rectangles)  as dumbbells of positive (white) and negative (black) magnetic charge (from \cite{Nisoli2010}). Optical microscopy of hexagonal (c) and square (d) SI, where the blue arrows denote the equivalent spins (from \cite{ortiz2016engineering}). Green disks show ice rule obeying vertices.}
\label{default}
\end{center}
\end{figure}

Indeed, despite  similarities, the two systems differ essentially in energetics and frustration. While local energetics promotes the ice rule in SI, it opposes it in PI. The energy of a SI vertex is typically proportional to the square of its topological charge, $E\propto q_n^2$, thus favoring the ice rule. For PI, it  is instead  $E \propto n(n-1)$, thus favoring  large negative charges which  violate the ice-rule (Figs.~2, 3).  In PI  the ice-manifold emerges as a collective energetic compromise in the thermodynamic limit~\cite{nisoli2014dumping,libal2017b} from the constraint that the total charge must be zero. It is thus locally unstable and fragile. It is a ``thin ice''. 

We  provide here a unifying framework  for the complex phenomenology of similarities and differences among the two classes of materials:  PI at equilibrium can be mapped directly into a SI coupled to a geometry-dependent background field.  In trivial geometries the field is zero and the two ices are equivalent. In non-trivial ones, however, it mediates the breakdown of the ice rule, leading to an entirely new phenomenology, still largely unexplored.  Without pretenses of exhaustiveness we propose some implications of this mapping to suggest exotic, novel behaviors which invite further experimental exploration. %We will suggest general effects by concentrating on particular cases. 

{\it 1---Isomorphism.} 
In PI, particles in positions $\{ {\bf y}\}$ repel with interaction $\phi(r)$. Clearly, their total energy
%
%\begin{equation}
${\cal H}= \sum_{{\bf y}\ne{\bf y'}}\phi\left(|{\bf y}-{\bf y'}|\right)$
%\label{H0}
%\end{equation}
%
does not appear very conducive to SI physics.
\begin{figure}[t!]
\center
\includegraphics[width=.83\columnwidth]{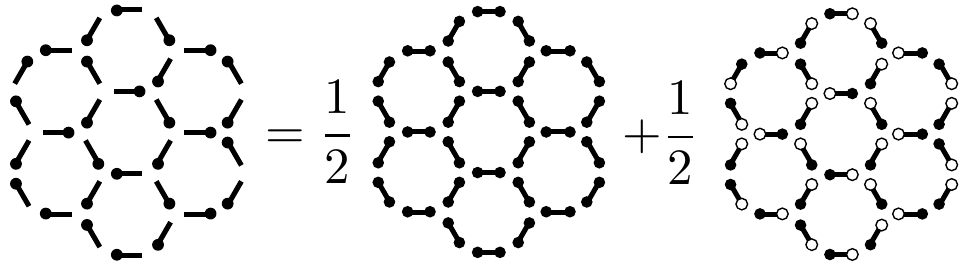}\vspace{2mm}
\includegraphics[width=.74\columnwidth]{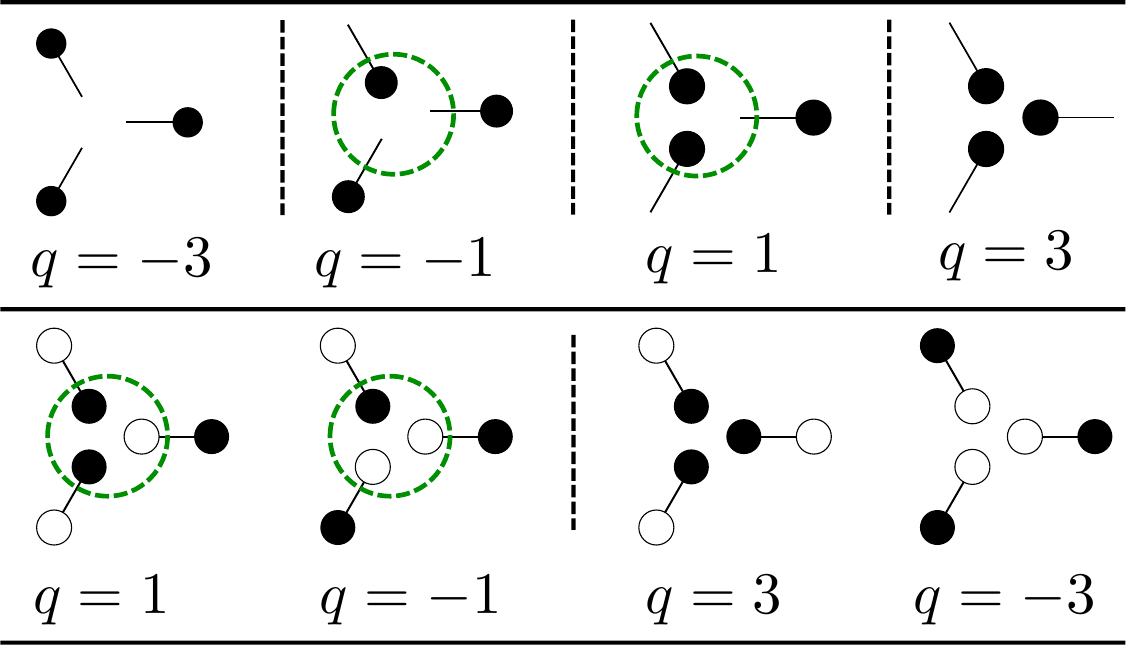}
\begin{center}
\caption{Top: Schematic illustration of Eq.~(\ref{iso}) where an hexagonal PI (here in a random configuration) is decomposed into a SI, with dipolar degrees of freedom, plus a background of positively saturated traps. The energy of the PI (middle) and SI (bottom) vertices, listed in increasing order  (from left to right, separated by dotted vertical lines)  differ essentially. PI promotes vertices of large negative charge, violating $\mathbb{Z}_2$ symmetry. SI promotes vertices of low absolute topological charge (ice rule, circled in green). }
\label{default}
\end{center}
\end{figure}
Yet, at equilibrium  the position of the particle in  a trap is a binary variable, represented by \textemdash\tikzcircle{3pt} or \tikzcircle{3pt}\textemdash   ~(Fig 1c,d). We can map PI into SI by ascribing a {\it positive} charge to our  \tikzcircle{3pt} particles, and introducing virtual {\it negative} charges \tikzcirclew{3pt}, which repel (resp. attract) other negative (positive) charges. Then, the energy does not change if we fractionalize each trap as a  trap doubly occupied by positive charges (or positive dumbbell  $\operatorname{ \tikzcircle{3pt}---\tikzcircle{3pt}}$), plus a dipole of negative and positive charges $\vec {\sigma} =  \operatorname{\tikzcirclew{3pt}---\tikzcircle{3pt}}$:
\begin{equation}
\operatorname{-----\tikzcircle{3pt}}=\frac{1}{2} \operatorname{ \tikzcircle{3pt}---\tikzcircle{3pt}}+\frac{1}{2} \operatorname{\tikzcirclew{3pt}---\tikzcircle{3pt}}.
\label{iso}
\end{equation}
Then  the energy  can    be rewritten  as
\begin{equation}
{\cal H}=\sum_{{\bf x}\ne { \bf x}'}V\left(\vec{\sigma}_{\bf x}, \vec{\sigma}_{\bf x'}\right)+\sum_{{\bf x}}W\left(\vec{\sigma}_{\bf x}\right)
%+\frac{1}{2}\sum_{\bf x}\left[\psi\left({\bf x}+\vec{\sigma}_{\bf x}/2\right) -\psi\left({\bf x-}\vec{\sigma}_{\bf x}/2\right)\right],
%+\sum_{\bf x}\vec{\sigma}_{\bf x} \cdot \vec{H}({\bf x})
\label{H}
\end{equation}
 (up to an irrelevant constant the  self-energy of the saturated traps). The first term represent the SI part of the hamiltonian: $V\left(\vec{\sigma}_{\bf x}, \vec{\sigma}_{\bf x'}\right)$ is the interaction between the dipolar spins $\vec{\sigma}_{\bf x}$ on the edges ${\bf x}$, and it can in general be reconstructed from  $\phi$. The second term is the interaction between dipoles and  positive dumbbells:
$W\left(\vec{\sigma}_{\bf x}\right)= [\psi\left({\bf x}+\vec{\sigma}_{\bf x}/2\right) -\psi\left({\bf x-}\vec{\sigma}_{\bf x}/2\right) ]/2,
%+\sum_{\bf x}\vec{\sigma}_{\bf x} \cdot \vec{H}({\bf x})
$ %
with
 $\psi({\bf x})=\sum_{{\bf y}_{\bullet}\ne {\bf x}}\phi ({\bf x}-{\bf y}_{\bullet})
$, where ${\bf y}_{\bullet}$ runs over all the allowed particle positions in all the dumbbells  $\operatorname{ \tikzcircle{3pt}---\tikzcircle{3pt}}$. 

Thus a PI is a SI under the  field $\psi$ generated by the virtual positive dumbbells. 
We will often adopt a nearest neighbor vertex model~\cite{Baxter1982,lieb1967residual} approximation and consider the ``energy of a vertex'', i.e. the interaction energies of all the spins impinging in said vertex. Figures~2, 3 show the different energy hierarchies for hexagonal and square geometries in the two pictures, where the SI picture recovers a $\mathbb{Z}_2$ symmetry absent in the PI picture. %While the two energetics  differ locally, they sum to the same value over the whole system. 

\begin{figure}[t!]
\center
\includegraphics[width=.9\columnwidth]{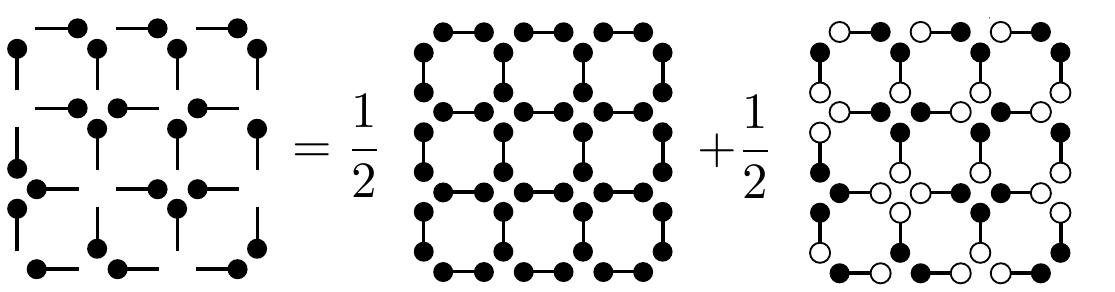}\vspace{2mm}
\includegraphics[width=.95\columnwidth]{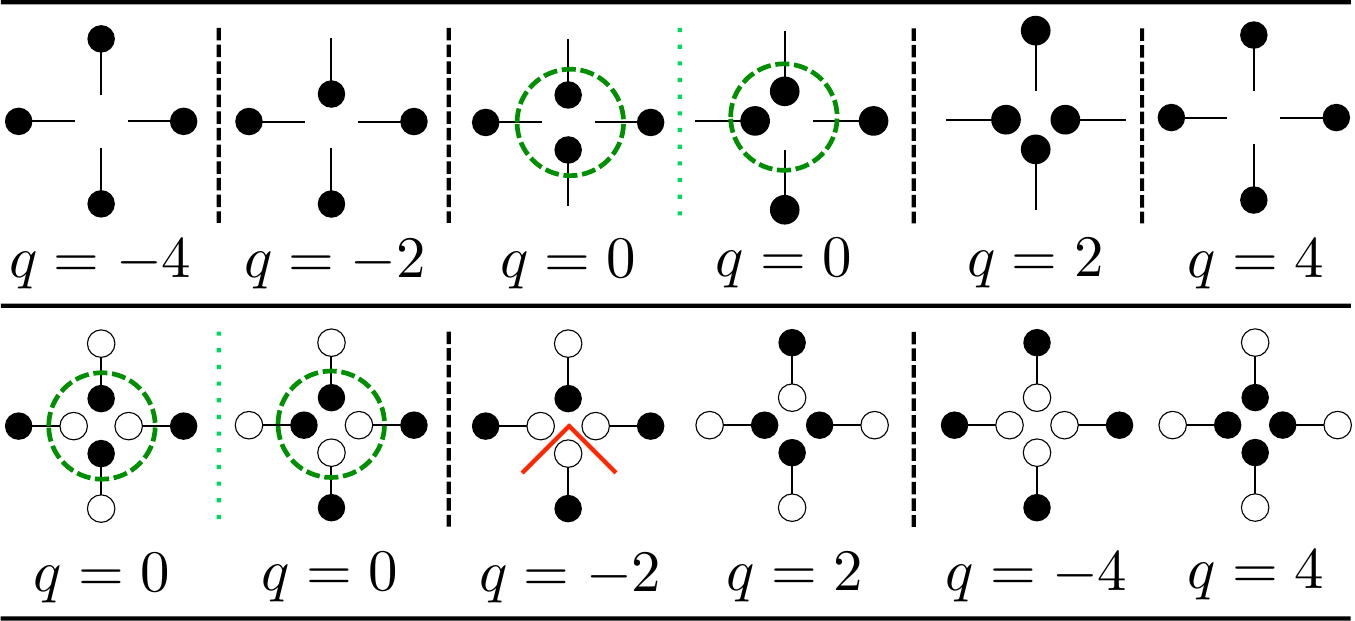}
\begin{center}
\caption{Same as in Fig.~2, but for the square lattice. Here, however, the degeneracy of the ice-rule vertices ($q=0$) is lifted by a difference in interaction strength between perpendicular and collinear traps or dumbbells (green dotted lines), and polarized  vertices have higher energy (middle: forth from left; bottom: second from left). The red  \textcolor{red}{$\wedge$} connects two plaquettes where the head-to-toe rule is broken  by a monopole (see  Fig.~5).}
\label{default}
\end{center}
\end{figure}
% 

%
%\begin{figure}[t!]
%\center
%\includegraphics[width=.95\columnwidth]{Fig2a.pdf}
%\begin{center}
%\caption{Schematic illustration of Eq.~(\ref{iso}). A particle ice can be decomposed into a dumbbell spin ice plus a background. Here the hexagonal geometry. }
%\label{default}
%\end{center}
%\end{figure}
%% 

We call a geometry {\it trivially-equivalent} if the second term in (\ref{H}) is a constant: then  the  the PI becomes a SI.  From symmetry considerations, ices whose vertices are the nodes of an infinite Bravais lattice are { \it trivially-equivalent}, explaining why the hexagonal and square PI  follow the ice rule~\cite{Libal2006,libal2009}.  
 
{\it 2---Ice rule and inner phases.} 
SI often exhibits layered phases. E.g.,  Kagome SI enters a charge-ordered/spin-disordered phase  within its  ice manifold, and then a long-range ordered, demagnetized phase within its  charge-ordered phase~\cite{Moller2009,Rougemaille2011,Chern2013,anghinolfi2015thermodynamic, Zhang2013}. A dipolar expansion of (\ref{H})
\begin{align}
{\cal H}&\simeq \frac{k}{2}\sum_{v} q^2_v + \frac{1}{2}\sum_{\langle v,v' \rangle} q_v q_{v'} \phi( r_{v,v'}) \nonumber \\
&+ \frac{1}{2} \sum_{{\bf x'}\notin \partial { \bf x}}{\sigma}^i_{\bf x} J_{ij}\left({\bf x}-{\bf x'}\right)  {\sigma}^j_{\bf x'}-\sum_{\bf x}\vec{\sigma}_{\bf x}\cdot \vec{E}({\bf x})
%+\sum_{\bf x}\vec{\sigma}_{\bf x} \cdot \vec{H}({\bf x})
\label{H2}
\end{align}
shows that PI also admits inner phases. The first term  imposes the ice rule  from the    interaction among dipoles within a vertex ($q_v$ is the charge of the vertex $v$,  and $k>0$ depends on $\psi$) and implies a crossover to an ice-manifold~\cite{Libal2006,libal2009}.  
The second term is an interaction between charged vertices and implies charge order at lower temperatures, as was recently seen numerically for PI with repulsion $\phi=x^{-3}$~\cite{libal2017inner}.
The third term is a generalized dipolar interaction among further neighboring spins  ($\partial {\bf x}$ is the neighborhood of ${\bf x}$) whose form depends on  $\phi$. For instance, for $\phi\propto r^{-\alpha}$ we have immediately
\begin{equation}
J_{ij}({\bf r})  \propto \left[\delta_{ij}-(\alpha+2)r_i r_j\right]{r^{-\alpha-2}},
\label{J}
\end{equation}
which reduces to the familiar dipolar interaction for {$\alpha=1$}. Instead,  $\alpha>1$ in (\ref{J})   strengthens the ferromagnetic term, leading to the ferromagnetic ordering within the disordered ice-manifold which has been recently obtained numerically in hexagonal PI~\cite{libal2017inner} for $\alpha=3$. %(Note that early works~\cite{Libal2006,libal2009} were based on particles  with short ranged repulsion of length-scale comparable to the lattice constant, and thus did not report any phase within the ice-manifold). 
In the fourth term of (\ref{H2}), $\vec E=-\vec \nabla \psi$ is the polarizing background field, whose role in ice rule fragility we will discuss now. %(on an infinite hexagonal or square lattice it is simply $\vec E=0$). 

{\it 3---Ice rule fragility: finite size systems.} Breakdown of the ice rule in PI follows from  its local energetics (Figs. 1, 2) lacking $\mathbb{Z}_2$ symmetry. Within the PI picture, it is explained as an effect of the background field in non-trivial geometries. 
One obvious case is a finite chunk of an otherwise trivially-equivalent structure. Then $\vec E$   comes from a finite chunk of positive dumbbells and points  toward the boundaries, polarizing  the spins outwards. The consequent accumulation of positive charges on the boundaries necessarily implies a violation of the ice rule in the bulk, as the net charge of all the dipoles is clearly zero. This polarization has been observed experimentally (reported to us  by P. Tierno, Barcelona). 

Instructively, this  ice rule break-down  disappears in the thermodynamic limit. The total negative charge in the system is proportional to the flux of $\vec{\sigma}_{\bf x}$ at the boundaries,  bound by their length $L$. Thus, the surface charge density goes to zero at least as $L^{-1}$: the ice-rule in PI is a collective effect only recovered in the thermodynamic limit. Nothing of the sort happens in SI~\cite{Li2010}. 

{\it 4) Ice rule fragility: mixed coordination.} More interesting is the ice rule breakdown in non-trivial, infinite lattices such as lattices of mixed coordination. Consider a regular or random decimation of a  trivially  equivalent structure such as the hexagonal ice. Because the background of {\it positive} dumbbells has no effect in the original  geometry, we can  express its effect in the decimated geometry as coming from {\it negative} virtual dumbbells, i.e. \textcolor{red}{{\tikzcircler{3pt}---\tikzcircler{3pt}}} traps saturated with negative particles,  placed in correspondence of the  decimated links (Fig.~4).

\begin{figure}[t!]
\center
\includegraphics[width=.75\columnwidth]{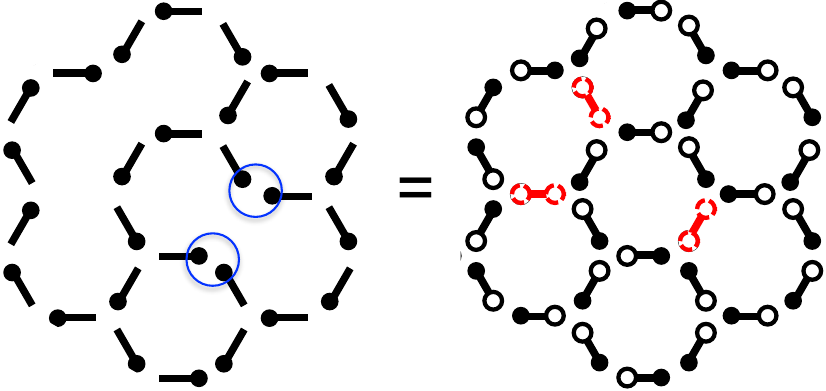}\vspace{2mm}
\includegraphics[width=.75\columnwidth]{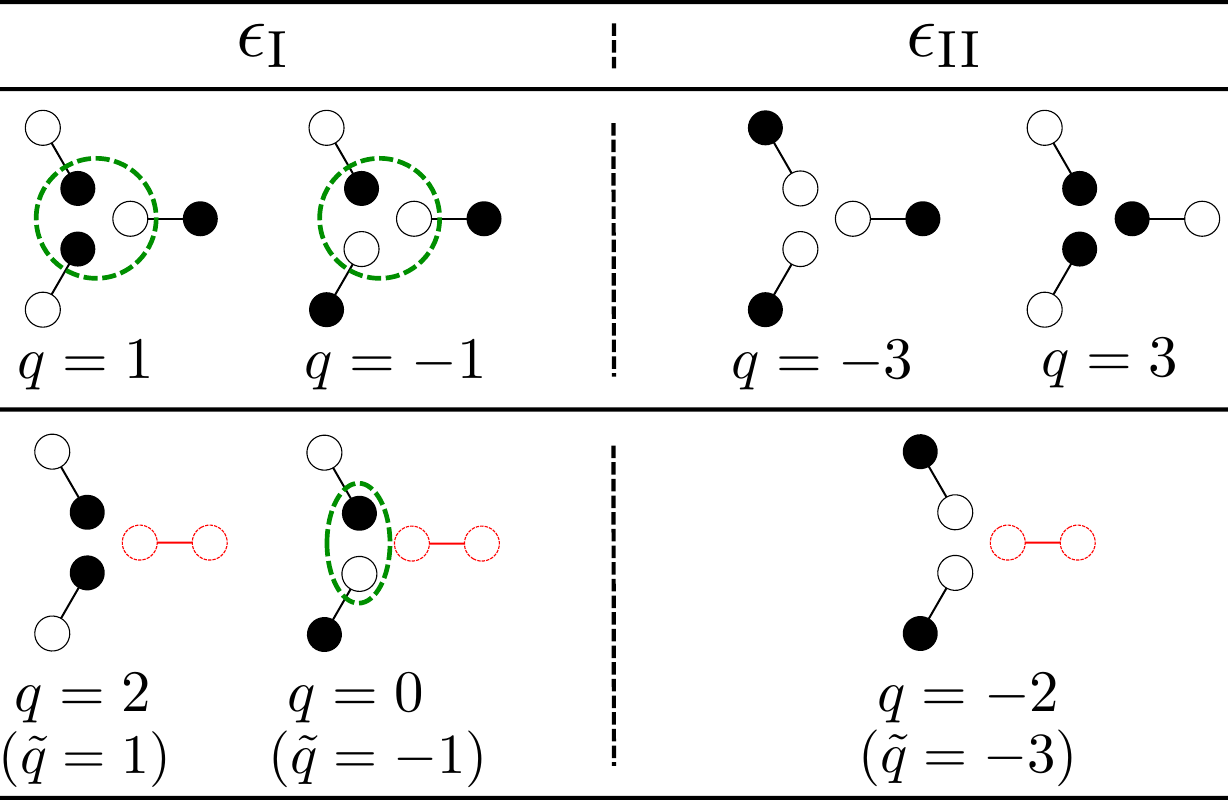}
\begin{center}
\caption{A portion of a decimated, infinitely extended PI (top left) in its low energy state can have ice rule violations on z=2 vertices ($q=+2$, blue circles) because it is equivalent to a  SI stuffed with virtual, negatively saturated traps (red, dashed) in lieu of the removed links (top right) with no violations of the ice rule in the virtual charge ($\tilde q=\pm1$). Indeed, while the SI energetics of the undecimated, $z=3$, vertices  (middle) is left unchanged, that of the decimated $z=2$ vertices (bottom) must include virtual negative saturated traps, and thus violates the ice rule at lowest energy:  the vertex of real charge $q=2$ (virtual charge $\tilde q=1$) is degenerate with the vertex of real charge $q=0$ (virtual charge $\tilde q=-1$), and with the $z=3$ vertices of real charge $q=\pm1$ (ice rule vertices circled in dashed green).}
\label{default}
\end{center}
\end{figure}

Crucially, in a  vertex-model approximation the energy of a vertex  is  proportional to its net {\it virtual} charge $\tilde q$, inclusive of the charge of  the negative, virtual dumbbell  \textcolor{red}{{\tikzcircler{3pt}---\tikzcircler{3pt}}}, breaking the $\mathbb{Z}_2$ symmetry of the SI energetics. As Fig.~4 shows, in  $z=2$ vertices the ice rule violating  2-in/0-out configuration of {\it virtual} charge $\tilde q=-1$ but {\it real}  positive charge $q=2$, has the same energy of the ice-rule configuration of 1-in/1-out ($q=0$, $\tilde q=+1$), but  also of the  ice rule  configurations of $z=3$ vertices. Thus,  $q=2$ charges appear entropically on  $z=2$ vertices in the degenerate ground state, in violation of the ice rule. The $z=3$ vertices remain in the ice rule, but    $q=-1$ charges must exceed $q=1$  ones, to cancel the positive charge on the $z=2$ vertices. 

This argument is completely general. Any mixed coordination lattice that can be obtained from decimating an ice-rule obeying PI must similarly show a transfer of topological charge from vertices of higher coordination to the decimated vertices of lower coordination, where charge is attracted by  negative virtual charges. Note that violation of the ice rule, however, do not necessarily happen in the lowest coordination vertices, as shown below and as found recently~\cite{libal2017b} in square PI. This charge transfer is unique to PI, and nothing of the sort can happen in SI. There the ice rule is robust to decimation and mixed coordination~\cite{Morrison2013,gilbert2014emergent,gilbert2016emergent}, dislocations~\cite{drisko2017topological}, and  indeed even in clusters~\cite{Li2010} as it is enforced by the local energy.

Finally, the charge transfer should be associated with glassiness. Indeed, while the average net charge must remain zero, or $q_{\mathrm{net}} \equiv N_v^{-1}\sum_v \langle q_v\rangle=0$ ($N_v$ is the number of vertices), its Edwards-Anderson parameter is not, or $q^2_{\mathrm{EA}} \equiv N_v^{-1} \sum_v \langle q_v\rangle^2\ne0$, because of the breakdown of the $\mathbb{Z}_2$ symmetry in the equivalent PI picture. This implies freezing of the charge in random distributions as $q^2_{\mathrm{EA}}$ is also the charge temporal autocorrelation function at large times.

{\it 5---Order breakdown from topological charge transfer.} It is well known that the ice manifolds of  square PI/SI are antiferromagnetically ordered because traps/spins  converging perpendicularly in the vertex interact more strongly than those converging  collinearly. This lifts the degeneracy of the ice-rule and favors the non-polarized, antiferromagnetic ice rule vertices~\cite{Wang2006,Porro2013, Zhang2013,Libal2006} of in Fig.~3. Consider a random decimations of traps in square PI that does not create $z=2$ vertices. It corresponds to a partial cover for a dimer cover model on the edges of the square lattice (Fig.~5a). In SI, such special decimation is  expected to preserve the antiferromagnetic order~\cite{Morrison2013}. In PI, instead, it  implies a structural transition to disorder, as shown below. 

Consider a decimation of the antiferromagnetic ensemble (Fig.~5b,c). Each virtual negative dumbbell  creates a negative  virtual monopole  $\tilde q=-2$ on a $z=3$ vertex. As explained above, the energy depends on the virtual charges. Because of (virtual and real) charge conservation, this decimated antiferromagnetically ordered state, which obeys the ice rule ($z=3$ vertices all have charges $q=\pm1$) has  the lowest energy. Is it  unique? 
\begin{figure}[t!]
\center
\includegraphics[width=1\columnwidth]{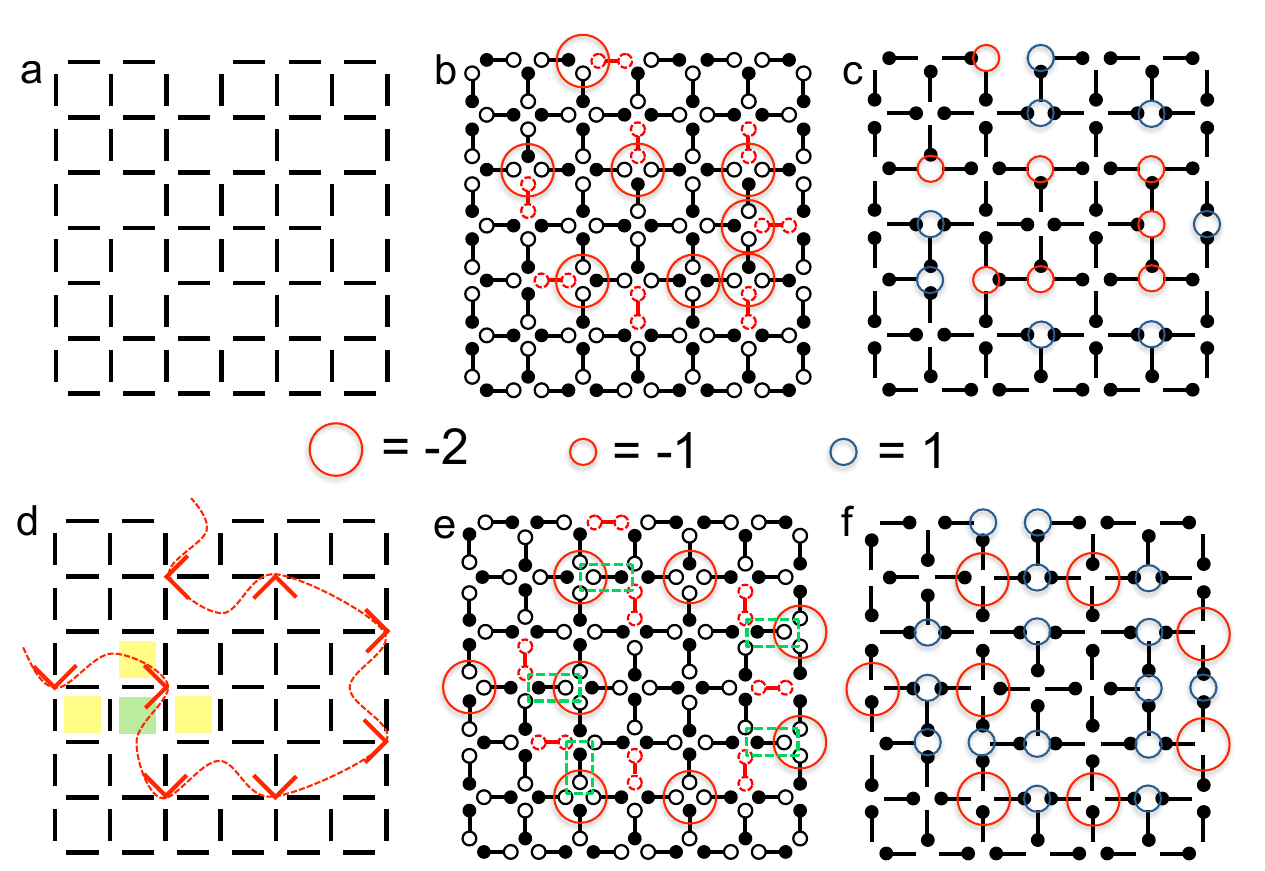}\vspace{2mm}
\begin{center}
\caption{Top (a-c): Decimating a square lattice (a)  {\it and} its anfiferromagnetic ground state  (decimated traps are replaced with negatively saturated traps, in red) leads to an ordered lowest energy state with $\tilde q=-2$ virtual charges on half of the decimated vertices in the SI picture (b). It corresponds to $q=\pm 1$ real charges on decimated vertices in the PI picture (c) and thus all vertices obey the ice rule. At low decimation, this is the only low energy state. Bottom (d-f):  However, above the decimation threshold corresponding to the percolation of decimated neighboring square plaquettes [e.g. yellow shaded ones neighboring a green one in (a)], the low energy state becomes degenerate. A disordered state can be chosen by connecting (red dotted line) neighboring decimated plaquettes (d) with $q=-2$ monopoles (represented with \textcolor{red}{$\wedge$} connectors as in Fig.~3)  on $z=4$ vertices, thus removing the virtual charges from decimated, $z=3$ vertices without increasing the energy [(e), green rectangles frame spins that can be freely flipped]. In the PI picture (f) this corresponds to ice rule violations on the $z=4$ vertices hosting  charge $q=2$: disorder comes from entropic transfer of topological charge.}
\label{default}
\end{center}
\end{figure}

At sufficiently low decimation it clearly is. Indeed, in a low energy state  only antiferromagnetic $q= 0$ ground state vertices,  and $q=-2$ monopoles are allowed on $z=4$ vertices. Therefore, in each minimal  plaquette the four dipoles must be arranged head to toe, except in correspondence of a monopole. In a plaquette affected by a virtual trap, a negative virtual charge sits on one of the two decimated vertices if and only if the remaining three spins are arranged head to toe. Antiferromagnetic order can break down if both virtual charges on the two neighboring decimated vertices are $\tilde q=0$ (and therefore their real charge is positive, $q=1$, on both). For that to happen, the head to toe rule must be broken on one of the two other vertices in each of the two relative plaquettes. It follows that two monopole of charge $q=-2$ must sit on the rectangular plaquette resulting from the decimation, as  monopoles are the only allowed vertices that can beak the head-to-toe rule. Because a monopole breaks the head-to-toe rule on two of the four plaquettes it separates (Fig.~3), this is only possible if at least one of the nearest neighboring plaquette is also decimated (Fig.~5d). Therefore, when the decimation is sufficiently low and the number of neighboring decimated plaquette is non-extensive the decimated antiferromagnetic state is the only ground state. However, when nearest neighboring decimated plaquette percolate (Fig.~5d), the low energy ensemble becomes disordered.

One can prove so by construction. Start with the decimated lattice, connect (or not) any neighboring decimated rectangular plaquette which can be connected via the red \textcolor{red}{$\wedge$}-connector  of Fig.~5, representing a monopole (Fig.~3). Then all the spins are determined. This construction corresponds to lines threading through decimated plaquettes (red dotted in Fig.~5c).  When the decimated plaquettes percolate at the nearest neighbor, these lines can be chosen freely either as closed loops or as infinite paths percolating through the material. This freedom in choosing connecting lines,  and more trivially the resulting free spins (Fig.~5e),  give a residual entropy to the ground state. There is thus a transition from order to a disordered state at a critical decimation,  likely a  glassy one involving dynamic arrest, and which invites experimental  analysis.

%
%If, however, another decimated link is in the proximity, it will also host a negative  then one can flip spins to remove both excitations
%
%The reason is that decimation of a $z=4$ lowest energy state does not create a $z=3$ lowest energy state (Fig. XXX).  because of the background field, here expressed through the virtual traps \tikzcirclew{3pt}---\tikzcirclew{3pt} impinging in the decimated vertices. When the decimation is sparse, each missing link creates two $z=3$, one in the ground state, one with the energy of a III$_4$ excitation (see Fig.~XXX). Relieving it via spin flip simply transports the excitations away, while creating  further excitations as II$_4$. If, however, another decimated link is in the proximity, then one can flip spins to remove both excitations, by creating n II$_4$ vertices, and changing the energy from $2\epsilon_{\mathrm{III}_4} \to n\epsilon_{\mathrm{II}_4}$. This process also leads to a charge increase on $z=3$ vertices.There is a maximum $n$ for which this process lowers the total energy, defined by the specific interaction $\phi$. Given $n$, when the decimation reaches a percolation threshold for the $n^{\mathrm{th}}$ neighbor, the system loses its order. Note that it can still be {\it locally} ordered, in absence of an antiferromagnetic order parameter, opening interesting perspective for a spin-ice to spin-glass transition at zero temperature. 
%
{\it Conclusion.} PI  can be considered a SI under a local fields that can break the SI's $\mathbb{Z}_2$ symmetry.  We have explored some of the implications and novel phenomenology which invite further numerical and experimental analysis. An extension to kinetics will be  explored in the future: as the current isomorphism starts from particles in their preferential binary locations  it does not capture their  motion across the trap, nor their intermediate interactions. Further developments  includes  the deliberate design of functional structural features, such as  interfaces which will accumulate positive charge and be therefore semipermeable to the passage of topologically charged defects under proper fields. 

 We wish to thank the LDRD office for financial support through  the ER program, A Libal, C. Reichhardt and CJ Olson Reichhardt (Los Alamos), and A. Ortiz and P. Tierno (Barcelona) for useful discussions and for sharing preliminary numerical and experimental results. This work was carried out under the auspices of the NNSA of the U.S. DoE at LANL under Contract No. DE-AC52-06NA25396.

\bibliography{library.bib}{}

\end{document}